\documentclass[conference]{IEEEtran}

\let\labelindent\relax

\usepackage{graphicx}
\usepackage[table,xcdraw]{xcolor}
\usepackage{tablefootnote}
\usepackage{url}
\usepackage{enumitem}
\usepackage{algorithm}
\usepackage{algpseudocode}

\begin{document}
\title{MFDPG: Multi-Factor Authenticated Password Management With Zero Stored Secrets}

\author{\IEEEauthorblockN{Vivek Nair}
\IEEEauthorblockA{UC Berkeley\\
vcn@berkeley.edu}
\and
\IEEEauthorblockN{Dawn Song}
\IEEEauthorblockA{UC Berkeley\\
dawnsong@berkeley.edu}
}

\IEEEoverridecommandlockouts
\makeatletter\def\@IEEEpubidpullup{6.5\baselineskip}\makeatother
\IEEEpubid{\parbox{\columnwidth}{
    Network and Distributed System Security (NDSS) Symposium 2024\\
    26 February - 1 March 2024, San Diego, CA, USA\\
    ISBN 1-891562-93-2\\
    https://dx.doi.org/10.14722/ndss.2024.23xxx\\
    www.ndss-symposium.org
}
\hspace{\columnsep}\makebox[\columnwidth]{}}

\maketitle

\begin{abstract}
While password managers are a vital tool for internet security, they can also create a massive central point of failure, as evidenced by several major recent data breaches.
For over 20 years, deterministic password generators (DPGs) have been proposed, and largely rejected, as a viable alternative to password management tools. In this paper, we survey 45 existing DPGs to asses the main security, privacy, and usability issues hindering their adoption. We then present a new multi-factor deterministic password generator (MFDPG) design that aims to address these shortcomings.
The result not only achieves strong, practical password management with zero credential storage, but also effectively serves as a progressive client-side upgrade of weak password-only websites to strong multi-factor authentication.
\end{abstract}

\section{Introduction}
Passwords constitute the primary authentication factor for the vast majority of currently deployed web applications. The prevalence of attacks like credential stuffing \cite{credential_stuffing} has made safe password management amongst the most critical security tasks that an average user faces today. Accordingly, a variety of popular password management tools have emerged to help users address this issue in a secure and convenient manner.

The most commonly used password management solutions are cloud-based applications like LastPass \cite{lastpass_arch}, Dashlane \cite{dashlane_arch}, and 1Password \cite{1password_arch}. These systems store user credentials in a centralized vault, typically encrypted using a key derived from the user's master password via PBKDF2 \cite{rfc2898}. In doing so, they afford users the convenience of accessing their accounts from any device, but also create a central point of failure that, if compromised, could reveal all of a user's passwords. Indeed, LastPass alone has experienced 8 major security incidents \cite{lastpass_security}, including a recent total data breach of stored credentials \cite{toubba_notice_2022}.

Alternatively, open-source password management software like Bitwarden \cite{bitwarden} and KeePassX \cite{keepassx} has emerged in part to eliminate the central point of failure created by cloud-based password managers. While popular amongst experienced users, these solutions have failed to achieve widespread adoption due their relative lack of usability. In particular, synchronizing data between devices is often a manual and cumbersome process.

\eject

For over two decades, deterministic password generators (DPGs) have been proposed as a substitute for all forms of password management \cite{karp2002site}. DPGs like PwdHash \cite{ross2005stronger} apply a cryptographic hash function to a user's master password and the domain name of a website to generate a unique pseudorandom password for each domain. As such, DPGs theoretically work seamlessly across devices without the need for synchronization, offering the security of a client-side password manager with the flexibility of a cloud-based solution. 

In practice, DPGs have been widely criticized for having a variety of security, privacy, and usability flaws \cite{noauthor_4_2016} that have thus far seriously hindered their adoption.
In this paper, we present a detailed analysis of 45 existing DPGs and summarize the main issues hindering their adoption.
In particular, current DPG schemes allow a user's master password to be directly attacked in the event that any of a user's generated passwords are compromised. Most DPGs also lack the flexibility to support password rotation or complex password policies.

Based on these findings, we present a new multi-factor deterministic password generator (MFDPG), which aims to rectify the shortcomings of existing DPGs by incorporating multi-factor key derivation \cite{mfkdf} into the password management process. In doing so, MFDPG effectively allows users to unilaterally upgrade password-only websites to support strong multi-factor authentication like TOTP \cite{rfc6238} and YubiKey \cite{yubikey}.
We further propose novel algorithmic solutions that facilitate password rotation and complex password policy compliance without leaking service usage patterns.
By presenting a truly secure and practical design, we hope to revive the notion of DPGs as a viable alternative to password management. \\

\noindent \textbf{Contributions:}
\vspace{-0.5em}
\begin{enumerate}[leftmargin=*]
    \item We analyze 45 existing DPGs to asses the security, privacy, and usability issues hindering their adoption (\S\ref{sec:survey}).
    \item We present MFDPG, a practical multi-factor authenticated DPG with zero client or server-side secret storage (\S\ref{sec:approach}).
    \begin{enumerate}
        \item MFDPG uses multi-factor key derivation to harden the system against attacks on the master password (\S\ref{sec:mfdpg}).
        \item We use Cuckoo filters to solve the revocation problem without leaking private account information (\S\ref{sec:revocation}).
        \item Our novel password generation algorithm supports any regular password policy using DFA traversal (\S\ref{sec:generation}).
    \end{enumerate}
    \item We evaluated MFDPG to verify its compatibility with all of the 100 most popular existing web applications (\S\ref{sec:evaluation}).
    \item MFDPG has the further effect of progressively upgrading any password-based website to support strong MFA (\S\ref{sec:discussion}).
\end{enumerate}

\eject
\section{Background and Related Work}

Passwords are notoriously weak as a sole authentication factor \cite{password_reuse, florencio_large_2006}, with attacks such as password spraying \cite{spraying} and credential stuffing \cite{credential_stuffing} affecting millions of online accounts each year. Nevertheless, a combination of circumstances has led passwords to remain the dominant authentication factor for online accounts today, making secure password management of paramount importance to one's overall security posture. An effective password management solution will allow users to have unique, high-entropy passwords for every website while maintaining security, portability, and ease of use.

In this section, we aim to motivate the need for a multi-factor deterministic password generator by discussing the range of existing password management and deterministic password generation solutions and their associated pitfalls.

\subsection{Password Managers}

The primary function of password managers is to provide an encrypted vault for the secure and portable storage of passwords. Today, users are faced with a difficult choice between the convenience of cloud-based password managers and the security of open-source self-managed solutions.

Cloud-based password managers, such as LastPass \cite{lastpass_arch}, Dashlane \cite{dashlane_arch}, and 1Password \cite{1password_arch}, are generally amongst the most popular password management solutions due to their perceived ease of use. The core technology used to secure these platforms is password-based key derivation.

Password-based key derivation functions like PBKDF2 \cite{rfc2898}, bcrypt \cite{bcrypt}, scrypt \cite{scrypt}, and Argon2 \cite{argon2} are one-way functions that convert a user's password into a cryptographic key that can be used for encryption.
Password-based key derivation functions are built upon cryptographic hash functions like SHA-256 \cite{rfc6234}, but also incorporate a degree of intentional computational inefficiency that increases the relative difficulty of brute-force attacks. For example, the PBKDF2 configuration used by LastPass uses 100,000 sequential rounds of SHA-256 to increase its computational difficulty.

Most cloud-based password managers use a password-based key derivation function such as PBKDF2 to derive a key from the user's password upon login.
A symmetric encryption function like AES-256 \cite{fips197} is then used to encrypt all of a user's secrets on the client side prior to their storage in a centralized database.
In theory, even an adversary with full access to the database will not be able to derive the key needed to decrypt the user's secrets without knowing their password.

Data breaches associated with major password managers are surprisingly common; LastPass alone has experienced at least 8 major security incidents \cite{lastpass_security}, including a recent total data breach of encrypted credentials \cite{toubba_notice_2022}. As no cloud-based service is totally immune from such a compromise, the security of these applications reduce, in practice, to the security of their users' master passwords. Given the aforementioned weakness of passwords as a sole authentication factor, there is a serious risk of attackers compromising the credentials stored in these services by performing offline brute-force attacks. As such, cloud-based password managers may even constitute a net liability for some users by creating a central point of failure.

Self-hosted open-source password managers like Bitwarden \cite{bitwarden}, KeePass \cite{reichl_keepass_nodate}, and KeePassX \cite{keepassx} use a similar architecture to cloud-based password managers to encrypt and store credentials using password-derived keys. However, these tools aim to address the vulnerability of cloud-based password managers to massive centralized data breaches by only storing encrypted passwords locally on a user's device, or on a server directly owned and maintained by the end user. 
In doing so, they avoid creating a concentrated high-value attack target as is inevitably the case with cloud-based password managers.

Still, self-hosted password managers are not immune to attack, with threats such as malware posing a risk to the database of encrypted credentials.
As with cloud-based password managers, the user's security in the event of a breach ultimately reduces to that of their master password.

Moreover, open-source password management solutions have failed to achieve widespread adoption due their perceived impracticality. The lack of a centralized database to store credentials requires users to either host and maintain their own servers, or to manually synchronize data between all of their devices. Further, while cloud-based password managers are usually designed around resilient and highly-available architectures, self-hosted solutions can be susceptible to a total loss of data if a single device is lost or rendered inoperable.
As such, most users have gravitated toward centralized password managers despite their relative security drawbacks.

\subsection{Deterministic Password Generators}
\label{sec:survey}

Deterministic password generators (DPGs) represent an interesting alternative to both cloud-based and self-hosted password management applications. Instead of storing encrypted passwords in any vault, DPGs work by deriving site-specific passwords using a cryptographic hash of the site's domain name and the user's master password. The deterministic nature of the underlying hash function ensures that the DPG will always generate the same password for a given site as long as the user remembers their master password, without needing to synchronize additional values between devices.
This basic idea was originally proposed by HP in 2002 \cite{karp2002site}, and was popularized by Stanford's PwdHash in 2003 \cite{PwdHash}. When implemented as a browser extension that automatically determines a site's URL, it has the additional benefit of resisting phishing attacks.

Since the advent of DPGs in the early 2000s, dozens of DPG tools have been developed and released as open-source websites, mobile applications, and browser extensions. Despite this, DPGs have seen even less mainstream adoption than self-hosted password managers, with the 20 most popular DPG extensions having less than 10,000 combined downloads.

Consumers' unwillingness to adopt DPGs may stem from a variety of security, privacy, and usability flaws \cite{noauthor_4_2016, palant_security_2016} that researchers have discovered with open-source DPGs.
However, the relative obscurity of DPGs as a mainstream tool has resulted in limited literature being generated on this topic.
To gain a better understanding of the current landscape of DPGs and their associated flaws, we performed a large-scale analysis of existing DPG implementations.
By searching GitHub, the Chrome Web Store, and the Firefox Extensions website, we identified 45 free software packages implementing some form of deterministic password generation algorithm.

We explored the source code of these 45 DPG extensions to identify whether there were any particular threats to security, privacy, or usability that may limit their use as an alternative to traditional password managers. Our findings for each DPG system are detailed in Table \ref{tab:dpg-survey}. Overall, we found four key issues affecting a large portion of the studied implementations:

\begin{enumerate}[leftmargin=*]
    \itemsep 1em
    \item \textbf{Brute-Force Susceptibility}. By far the largest concern with current DPG implementations is the ability to attack a user's master password. Because of the lack of stored ciphertexts in DPGs, exposure of the master password is sufficient to reveal all of a user's site passwords.    
    However, as it stands, obtaining any of a user's site-specific passwords allows an attacker to perform an offline brute-force attack on that user's master password by checking which master password would, when combined with the site's known URL, have resulted in the correct site password being generated.
    
    \medskip
    
    This threat would be somewhat mitigated by the use of a strong progressive password-based key derivation function to increase the computational difficulty of a brute-force attack. Unfortunately, of the 45 applications surveyed, 28 only used a standard cryptographic hash, with 11 using deprecated function (MD5 or SHA1). A further 13 applications did use a progressive hash function, but supplied a cost parameter too low to effectively deter modern hardware. Only four of the 45 applications implemented a progressive hash function with well-chosen cost parameters.
    
    \item \textbf{No Policy Support}. A second issue that affects most DPGs is the lack of support for applications with complex password policies.
    Websites often enforce a number of password strength and complexity requirements, with length, capitalization, character sets, etc. varying greatly from one service to another. No single static generator can hope to accomodate a wide range password policies without tuning the generation algorithm for each site.
    
    \medskip
    
    Of the 45 DPGs surveyed, 27 provide no settings at all for customizing password generation; users of these applications would likely face scenarios in which the DPG cannot generate a password compatible with a service they wish to use. The remaining 18 provide limited customization, which in most cases only allows the output length to be tweaked. Users of these applications must also remember these settings and configure them correctly on every login.
    
    \item \textbf{No Revocation Support}. Similarly, websites often impose password rotation policies that require users to periodically change their password. However, the deterministic nature of DPGs usually does not allow users to have more than one password for a given website. Only four of the DPGs we analyzed provided a way to revoke a password and generate a new one for the same service. Of these, two required users to remember a revocation counter value for each website, while the other two stored this counter value in a file. 
    
    \item \textbf{Multi-Factor Authentication}. Finally, a fundamental limitation of all existing DPGs is the solitary reliance on a master password and the lack of support for multi-factor authentication. Most centralized password managers are enhanced by their support for multi-factor authentication at the login stage, though password vaults are still ultimately only encrypted with password-derived keys.
\end{enumerate}

In addition to the common issues noted above, we found specific vulnerabilities in a six implementations: two schemes designed their own insecure hash function, two incorrectly deployed progressive hash functions, and two had obvious network-related vulnerabilities. More information about these issues is given in the footnotes of Table \ref{tab:dpg-survey}.

As a result of many of the discussed issues, DPGs have remained a relatively obscure technology and have largely been dismissed as a viable alternative to conventional password managers. Thus, DPG projects have largely been abandoned, with most open-source DPGs receiving no updates in the past five years, as illustrated in Table \ref{tab:dpg-survey}. We hope, in this paper, to revive the field of deterministic password generation, aided by recent cryptographic advances that enable the incorporation of multiple authentication factors into derived keys.

\subsection{Multi-Factor Authentication}

The most popular form of multi-factor authentication in use today is ``out-of-band authentication'' (OOBA) \cite{ooba}, which includes email and SMS-based authentication. However, these channels are not trustless and fundamentally require a server; thus, they do not receive significant attention in this paper.

Aside from OOBA, popular MFA options include ``soft tokens'' like HMAC-based one-time password (HOTP) \cite{rfc4226} and time-based one-time password (TOTP) \cite{rfc6238}, as implemented in applications like Google Authenticator. These factors use a counter or timestamp to generate one-time passwords based on a pre-shared secret.
``Hard tokens'' such as YubiKeys \cite{yubikey} are also a popular option that requires using specialized hardware.

\subsection{Multi-Factor Key Derivation}
The Multi-Factor Key Derivation Function (MFKDF) \cite{mfkdf} is a recent improvement over PBKDFs that incorporates multiple authentication factors into the key derivation process. It is a trustless cryptographic operation that can handle many popular authentication factors like HOTP, TOTP, and YubiKeys on the client side without the need for a trusted server.

The MFKDF specification contains two major architectural components. The first component is the set of so-called ``factor constructions,'' which convert a dynamic \textit{factor witness}\footnote{In this case, the \textit{witness} refers to the message used to authenticate (e.g., a 6-digit OTP), which is often not the same as the underlying shared secret.} and public parameters into static key material. The public parameters require no security assumptions and can safely be stored in a database without concern for revealing information about the factors to potential adversaries. Constructions are given for a variety of popular authentication factors.

The second major component of MFKDF is the key derivation function itself, which adds a secret sharing layer to provide functionality such as threshold-based key derivation, advanced policy enforcement, and factor recovery. Together, these components convert multiple authentication factors into a cryptographic key, and serve as a replacement for PBKDFs. 

This paper proposes using MFKDF to enhance the security and utility of classical DPGs. While DPGs have been around in some form since 2002, we believe the recent introduction of MFKDF has provided a critical tool for their practical use.

\eject
\onecolumn
\begin{table}[h]
\centering
\renewcommand{\arraystretch}{1.01}%
\resizebox{\textwidth}{!}{%
\begin{tabular}{|l|l|l|l|l|l|}
\hline
\textbf{Name} & \textbf{Last Updated} & \textbf{Hash Function (Cost)} & \textbf{Policies?} & \textbf{Revocation?} & \textbf{Flaws?} \\ \hline
PasswordMaker \cite{PasswordMaker} & 08/2010 & {\color[HTML]{FF0000} SHA1/MD5} & {\color[HTML]{F1C232} Limited} & {\color[HTML]{FF0000} No} &  \\ \hline
PasswordProtect \cite{PasswordProtect} & 07/2011 & {\color[HTML]{F1C232} PBKDF2-SHA1 (10000)} & {\color[HTML]{FF0000} No} & {\color[HTML]{FF0000} No} &  \\ \hline
Password Hasher \cite{Password_Hasher} & 01/2012 & {\color[HTML]{FF0000} SHA1} & {\color[HTML]{F1C232} Limited} & {\color[HTML]{FF0000} No} &  \\ \hline
Vault \cite{Vault} & 07/2012 & {\color[HTML]{F1C232} PBKDF2-SHA1 (8)} & {\color[HTML]{F1C232} Limited} & {\color[HTML]{FF0000} No} &  \\ \hline
RndPhrase \cite{RndPhrase} & 10/2012 & {\color[HTML]{FF0000} CubeHash} & {\color[HTML]{FF0000} No} & {\color[HTML]{FF0000} No} &  \\ \hline
Magic \cite{Magic} & 12/2012 & {\color[HTML]{FF0000} Custom} & {\color[HTML]{FF0000} No} & {\color[HTML]{FF0000} No} & {\color[HTML]{FF0000} Yes\tablefootnote{Magic Password Generator uses a custom hash function that is seemingly not a true one-way function.}} \\ \hline
BPasswd \cite{BPasswd} & 01/2013 & {\color[HTML]{F1C232} bcrypt (64)} & {\color[HTML]{FF0000} No} & {\color[HTML]{FF0000} No} &  \\ \hline
My Password \cite{My_Password} & 06/2013 & {\color[HTML]{FF0000} MD5} & {\color[HTML]{FF0000} No} & {\color[HTML]{FF0000} No} &  \\ \hline
Hash Password \cite{Hash_Password} & 07/2014 & {\color[HTML]{FF0000} Custom} & {\color[HTML]{F1C232} Limited} & {\color[HTML]{FF0000} No} & {\color[HTML]{FF0000} Yes\tablefootnote{Hash Password Generator uses a custom hash function that is seemingly not a true one-way function.}} \\ \hline
SecPassGen \cite{SecPassGen} & 08/2014 & {\color[HTML]{F1C232} PBKDF2-SHA1 (10000)} & {\color[HTML]{FF0000} No} & {\color[HTML]{FF0000} No} &  \\ \hline
pastor \cite{pastor} & 08/2014 & {\color[HTML]{F1C232} PBKDF2-SHA256 (1000)} & {\color[HTML]{FF0000} No} & {\color[HTML]{FF0000} No} &   \\ \hline
Passera \cite{Passera} & 09/2014 & {\color[HTML]{FF0000} SHA512} & {\color[HTML]{F1C232} Limited} & {\color[HTML]{FF0000} No} &  \\ \hline
pwgen \cite{pwgen} & 10/2014 & {\color[HTML]{FF0000} MD5} & {\color[HTML]{FF0000} No} & {\color[HTML]{FF0000} No} &  \\ \hline
PswGen Toolbar \cite{PswGen_Toolbar} & 11/2014 & {\color[HTML]{FF0000} SHA512} & {\color[HTML]{FF0000} No} & {\color[HTML]{FF0000} No} &  \\ \hline
Extrasafe \cite{Extrasafe} & 04/2015 & {\color[HTML]{FF0000} SHA3} & {\color[HTML]{F1C232} Limited} & {\color[HTML]{FF0000} No} &  \\ \hline
determ-pwgen \cite{determ-pwgen} & 05/2015 & {\color[HTML]{FF0000} SHA512} & {\color[HTML]{F1C232} Limited} & {\color[HTML]{FF0000} No} &  \\ \hline
HashPass \cite{HashPass} & 06/2015 & {\color[HTML]{FF0000} SHA1/MD5} & {\color[HTML]{FF0000} No} & {\color[HTML]{FF0000} No} &  \\ \hline
hash0 \cite{hash0} & 07/2015 & {\color[HTML]{6AA84F} PBKDF2-SHA256 (100000)} & {\color[HTML]{FF0000} No} & {\color[HTML]{FF0000} No} &  \\ \hline
vPass \cite{vPass} & 08/2015 & {\color[HTML]{FF0000} TEA (10)} & {\color[HTML]{FF0000} No} & {\color[HTML]{FF0000} No} &  \\ \hline
Recall my password \cite{Recall_my_password} & 09/2015 & {\color[HTML]{FF0000} SHA512} & {\color[HTML]{FF0000} No} & {\color[HTML]{FF0000} No} & {\color[HTML]{FF0000} Yes\tablefootnote{Recall my password sends the user's password to the author's website to perform the hashing on the backend.}} \\ \hline
MS \cite{MS} & 10/2015 & {\color[HTML]{FF0000} SHA1} & {\color[HTML]{F1C232} Limited} & {\color[HTML]{FF0000} No} &  \\ \hline
BPasswd2 \cite{BPasswd2} & 11/2015 & {\color[HTML]{F1C232} bcrypt (64)} & {\color[HTML]{F1C232} Limited} & {\color[HTML]{FF0000} No} &  \\ \hline
Domain \cite{Domain} & 01/2016 & {\color[HTML]{FF0000} SHA1} & {\color[HTML]{FF0000} No} & {\color[HTML]{FF0000} No} &  \\ \hline
Password Maker X \cite{Password_Maker_X} & 02/2016 & {\color[HTML]{FF0000} SHA1/MD5} & {\color[HTML]{FF0000} No} & {\color[HTML]{FF0000} No} &  \\ \hline
UniPass \cite{UniPass} & 03/2016 & {\color[HTML]{F1C232} PBKDF2-SHA256 (4096)} & {\color[HTML]{FF0000} No} & {\color[HTML]{FF0000} No} &  \\ \hline
passwordgen \cite{passwordgen} & 04/2016 & {\color[HTML]{FF0000} SHA256} & {\color[HTML]{FF0000} No} & {\color[HTML]{FF0000} No} &  \\ \hline
uPassword \cite{uPassword} & 05/2016 & {\color[HTML]{FF0000} SHA1} & {\color[HTML]{F1C232} Limited} & {\color[HTML]{F1C232} Local} &  \\ \hline
Pegasus \cite{Pegasus} & 11/2016 & {\color[HTML]{6AA84F} PBKDF2-SHA512 (555000)} & {\color[HTML]{F1C232} Limited} & {\color[HTML]{F1C232} Manual} & {\color[HTML]{FF0000} Yes\tablefootnote{The PBKDF2 hash only uses the master password. That key is then combined with the site name via SHA512. This makes breaking PBKDF2 unnecessary.}} \\ \hline
strongpass \cite{strongpass} & 11/2016 & {\color[HTML]{F1C232} scrypt (16384)} & {\color[HTML]{FF0000} No} & {\color[HTML]{F1C232} Manual} &  \\ \hline
Persistent Generator \cite{Persistent_Generator} & 03/2017 & {\color[HTML]{FF0000} MurmurHash} & {\color[HTML]{F1C232} Limited} & {\color[HTML]{FF0000} No} &  \\ \hline
PwdHash \cite{PwdHash} & 05/2017 & {\color[HTML]{FF0000} HMAC-MD5} & {\color[HTML]{FF0000} No} & {\color[HTML]{FF0000} No} & {\color[HTML]{FF0000} Yes\tablefootnote{PwdHash requires you to enter the master password into a web page that loads external scripts and contains a basic XSS vulnerability.}} \\ \hline
Phashword \cite{Phashword} & 10/2017 & {\color[HTML]{FF0000} SHA1} & {\color[HTML]{F1C232} Limited} & {\color[HTML]{FF0000} No} &  \\ \hline
python-dpg \cite{python-dpg} & 10/2017 & {\color[HTML]{FF0000} SHA256} & {\color[HTML]{FF0000} No} & {\color[HTML]{FF0000} No} &  \\ \hline
LastWord \cite{LastWord} & 12/2017 & {\color[HTML]{FF0000} SHA1} & {\color[HTML]{FF0000} No} & {\color[HTML]{FF0000} No} &  \\ \hline
MasterPassX \cite{MasterPassX} & 02/2018 & {\color[HTML]{FF0000} HMAC-SHA256} & {\color[HTML]{FF0000} No} & {\color[HTML]{FF0000} No} &  \\ \hline
Art \cite{Art} & 03/2018 & {\color[HTML]{F1C232} PBKDF2-SHA256 (100)} & {\color[HTML]{FF0000} No} & {\color[HTML]{FF0000} No} &  \\ \hline
Tresor \cite{Tresor} & 12/2018 & {\color[HTML]{F1C232} PBKDF2-SHA1 (8)} & {\color[HTML]{F1C232} Limited} & {\color[HTML]{FF0000} No} &  \\ \hline
mypass \cite{mypass} & 02/2019 & {\color[HTML]{FF0000} Skein} & {\color[HTML]{F1C232} Limited} & {\color[HTML]{F1C232} Local} &  \\ \hline
PasswordShaker \cite{PasswordShaker} & 05/2019 & {\color[HTML]{6AA84F} scrypt (32768)} & {\color[HTML]{F1C232} Limited} & {\color[HTML]{FF0000} No} &  \\ \hline
Aprico \cite{Aprico} & 06/2019 & {\color[HTML]{F1C232} scrypt (16384)} & {\color[HTML]{FF0000} No} & {\color[HTML]{FF0000} No} &  \\ \hline
PasswordBuilder \cite{PasswordBuilder} & 06/2019 & {\color[HTML]{F1C232} scrypt (1024)} & {\color[HTML]{FF0000} No} & {\color[HTML]{FF0000} No} &  \\ \hline
CCTOO \cite{CCTOO} & 09/2019 & {\color[HTML]{F1C232} scrypt (16384)} & {\color[HTML]{FF0000} No} & {\color[HTML]{FF0000} No} &  \\ \hline
Hasher Plus \cite{Hasher_Plus} & 10/2019 & {\color[HTML]{FF0000} HMAC-SHA1} & {\color[HTML]{F1C232} Limited} & {\color[HTML]{FF0000} No} &  \\ \hline
masterpassword \cite{masterpassword} & 03/2022 & {\color[HTML]{6AA84F} scrypt (32768)} & {\color[HTML]{FF0000} No} & {\color[HTML]{FF0000} No} & {\color[HTML]{FF0000} Yes\tablefootnote{The scrypt hash only uses the username and master password. That key is then combined with the URL via SHA256. This makes breaking scrypt unnecessary.}} \\ \hline
Passcrambler \cite{Passcrambler} & 01/2023 & {\color[HTML]{FF0000} MD5} & {\color[HTML]{F1C232} Limited} & {\color[HTML]{FF0000} No} &  \\ \hline
\end{tabular}%
}
\vspace{0.5em}
\caption{Survey of 45 deterministic password generation tools and extensions.}
\vspace{-2.5em}
\label{tab:dpg-survey}
\end{table}
\twocolumn

\section{Problem Statement}
\label{sec:problem}

Our goal is to present a deterministic password generator (DPG) design that remedies the security, privacy, and usability issues of existing DPGs. In this section, we detail the specific goals and assumptions of the presented MFDPG construction.

\subsection{Threat Model}
Consider a DPG implemented as a client-side application or browser extension with optional server-side storage of settings, salts, and other trustless public parameters. 
We consider security under a \textit{total data breach} threat model; i.e., at some instant, an adversary receives a snapshot of all materials stored on the client or server. The adversary may use all such data to attempt to violate one or more of the below security goals.



\subsection{Security Goals}
\label{sec:goals}

Our design must first and foremost satisfy the following standard properties of DPGs. Though existing DPGs use a master password as a sole input factor, we generalize these definitions to support a collection of input factors for forward-compatibility with our scheme. Let one such set of input factors (and corresponding output) be defined as ``correct'' for a given user. The desired properties are then as follows:

\begin{enumerate}[leftmargin=*]
    \itemsep 1em
    \item \textbf{Correctness} (Determinism). Given a static set of ``correct'' input factors, the scheme always outputs the same ``correct'' password for any target service; i.e., it is deterministic. 
    \item \textbf{Safety} (Pseudorandomness). Given an ``incorrect'' set of input factors, the scheme outputs an ``incorrect'' password for any target service except with negligible probability. 
    \item \textbf{Secretless}. While password managers can achieve the above properties through stateful storage of ciphertexts, a DPG should not store site-specific ciphertexts in any location, i.e., it is stateless other than public parameters.
\end{enumerate}

Because the goal of this paper is to significantly improve upon the current state-of-the-art DPGs, we impose a few additional requirements on our MFDPG scheme:

\begin{enumerate}[resume,leftmargin=*]
    \itemsep 1em
    \item \textbf{Brute-Force Resistance}. A brute-force attack on the master password should be insufficient to generate site-specific passwords. Adversaries that obtain a site-specific password should not easily be able to attack the master password.
    \item \textbf{Compatibility}. The DPG should be compatible with a wide variety of password policies. Namely, the DPG should be able to generate passwords compliant with any regular password policy (i.e., representable by a regular expression).
    \item \textbf{Revocability}. Users should, at any time, be able to revoke a password for a given service and generate a new one for the same service, without changing their input factors, and without manually remembering a counter value.
    \item \textbf{Privacy}. In implementing the above properties, a DPG should not store site-specific values that allow any outside adversary to identify the services utilized by the DPG user.
\end{enumerate}

\eject
\section{Proposed Approach}
\label{sec:approach}

We now present our proposed design for a multi-factor deterministic password generator (MFDPG). We begin with an overview of our construction and then specifically address the issues of password policy support and revocation in dedicated sections. Our design places a focus on modularity, with any of these components being replaceable in future iterations.

\subsection{Multi-Factor Password Generation}
\label{sec:mfdpg}

\begin{figure}[H]
\centering
\includegraphics[width=\linewidth]{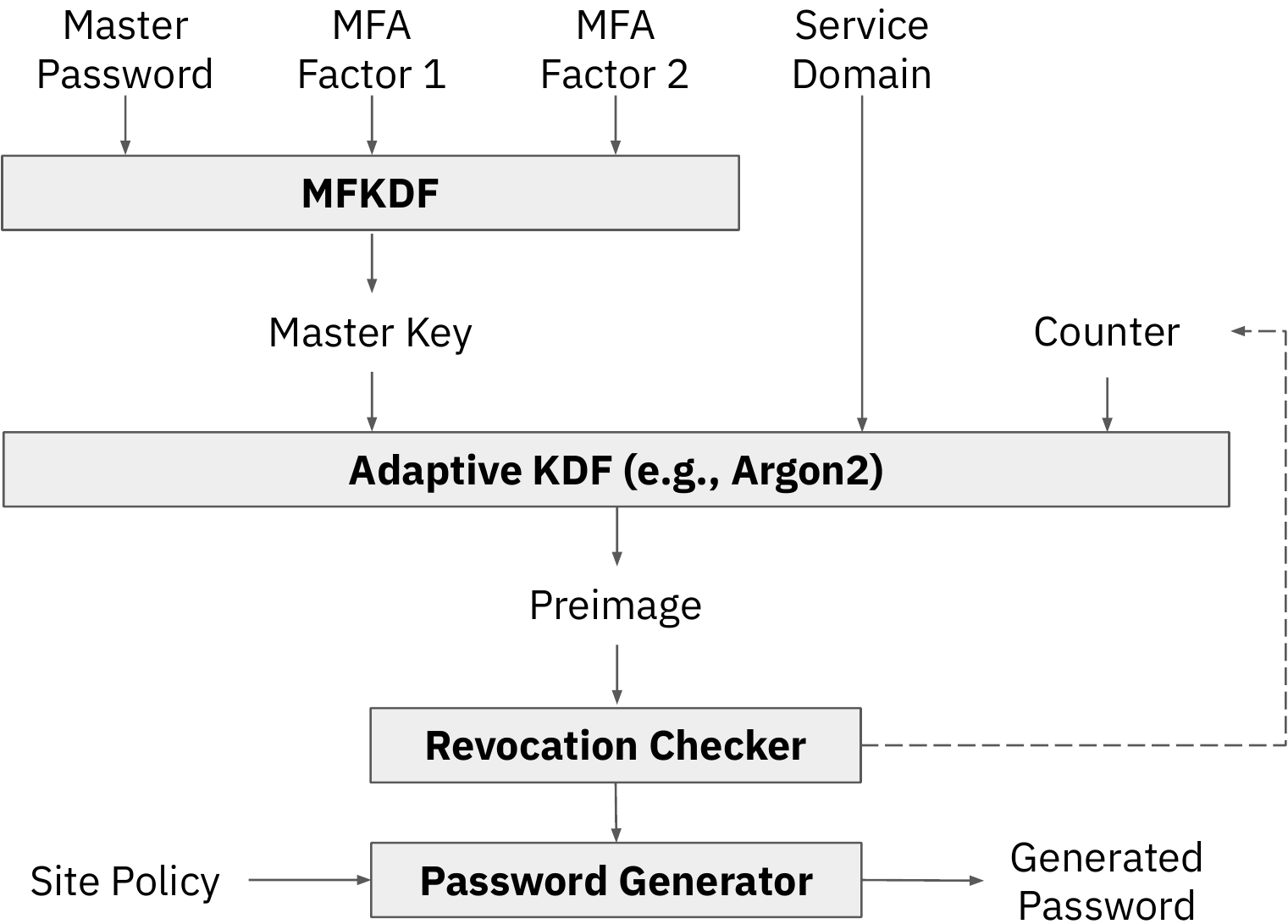}
\caption{Overview of multi-factor password generation architecture.}
\label{fig:mfdpg}
\end{figure}

Fig. \ref{fig:mfdpg} shows an overview of our general approach for deriving a site-specific password from multiple authentication factors. The derivation algorithm (Alg. \ref{alg:mfdpg}) follows 4 general steps (the detailed algorithms for revocation and generation in steps 3 and 4 are discussed in \S\ref{sec:revocation} and \S\ref{sec:generation}, respectively):

\begin{enumerate}[leftmargin=*]
    \item Use MFKDF to derive a master key from multiple factors.
    \item Use an adaptive KDF to derive a site-specific preimage.
    \item Check if the preimage is revoked, and iterate until clear.
    \item Convert preimage into a site policy-compliant password.
\end{enumerate}

The most obvious difference with this approach in comparison with traditional DPGs is the use of MFKDF to support multiple authentication factors, including YubiKeys, HOTP, TOTP, etc., in addition to passwords. In some cases, doing so requires the storage of public parameters (e.g., salts or one-time pads) either locally or in the cloud. Importantly, these values are entirely trustless; for example, MFKDF-based cryptocurrency wallets \cite{mfkdf_wallet} store them openly on blockchains.

\begin{algorithm}[H]
\caption{MFDPG Algorithm}
\label{alg:mfdpg}
\begin{algorithmic}[1]
\Require $\mathsf{M}$ is MFKDF per Nair and Song \cite{mfkdf}
\Require $\mathsf{K}$ is an adaptive KDF (e.g., Argon2 \cite{argon2})

\Function{MFDPG}{$\mathsf{factors}, \mathsf{service}, \mathsf{policy}$}
    \State $\mathsf{master\_key} \gets \mathsf{M}(\mathsf{factors})$
    \State $\mathsf{counter} \gets 0$
    \Repeat
        \State $\mathsf{counter} \gets \mathsf{counter} + 1$
        \State $\mathsf{preimage} \gets \mathsf{K}(\mathsf{master\_key} \odot \mathsf{service} \odot \mathsf{counter})$
    \Until \textbf{not} {$\textsc{Check}(\mathsf{preimage})$}
    \State \Return $\textsc{Generate}(\mathsf{preimage}, \mathsf{policy})$
\EndFunction

\end{algorithmic}
\end{algorithm}

Using MFKDF rather than a PBKDF addresses the most significant issue with current PRGs: brute-force attack susceptibility. Because of the ``exponential entropy'' property of MFKDF, adversaries can no longer use a site-specific password to attack the user's master password. Instead, they would have to simultaneously crack all of a user's factors, a task that is significantly harder than guessing their master password alone.

An additional advantage of using MFKDF in this context is its support for factor recovery. MFKDF presents a threshold variant that allows 3 authentication factors to be established, any 2 of which can be used to derive the key. When using this variant, the resulting MFDPG construction could allow users to recover from the loss of a single authentication factor without losing access to all of their generated passwords, as is the case in present password-only DPGs.

\subsection{Revocation Algorithm}
\label{sec:revocation}

Another key difficulty with current PRGs is the lack of revocation support. As discussed in \S\ref{sec:survey}, websites often implement password rotation policies that require users to update their password occasionally. Our goal is to support revocation in a way that does not store information about the services a user has accounts on, but also does not require users to remember and manually enter a unique revocation counter value for each service. To achieve this, we suggest using a Cuckoo filter \cite{fan2014cuckoo} or a similar set membership data structure.

\begin{figure}[H]
\centering
\includegraphics[width=0.75 \linewidth]{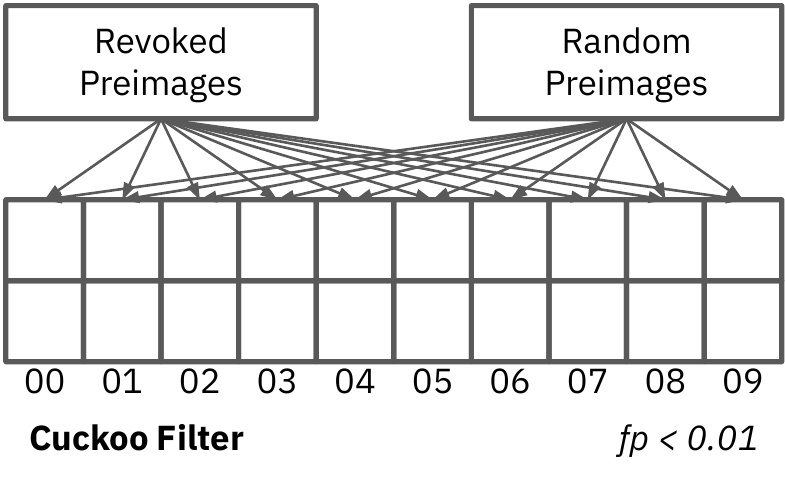}
\caption{Fixed-capacity Cuckoo filter for private credential revocation support.}
\label{fig:cuckoo}
\end{figure}

Our suggested approach, illustrated in Fig. \ref{fig:cuckoo}, involves filling a Cuckoo filter with both true, revoked preimages and deterministic fictitious preimages, such that the total number of entries remains static. These true and fictitious entries should be indistinguishable, such that the number of true revoked items in the set should not be discernable to an adversary with access to the Cuckoo filter but not the user's master key. 

In addition to its time and space efficiency, the use of a probabilistic set membership data structure has the advantage of not revealing its exact members.
When correctly tuned, this approach is likely to further frustrate brute-force attacks, with a high false-positive rate for attackers, while still posing sufficiently low friction for users. As a result, users can revoke credentials as needed. Unlike current DPGs, they may do so without storing information that reveals the services they use, or even the number of revocations that they have performed. The exact revocation algorithm is detailed in Alg. \ref{alg:revocation}.

\begin{algorithm}[H]
\caption{MFDPG Revocation Algorithms}
\label{alg:revocation}
\begin{algorithmic}[1]
\Require $\mathsf{S}$ is a Cuckoo filter per Fan et al. \cite{fan2014cuckoo}
\Require $\mathsf{H}$ is a cryptographic hash function (e.g., SHA256)
\Require $\mathsf{N}$ is the maximum number of allowed revocations

\Function{Setup}{$\mathsf{master\_key}$}
    \For{$\mathsf{i}~\textbf{in}~[0\ldots\mathsf{N}]$}
        \State $\mathsf{S}.\mathsf{add}(H(\mathsf{master\_key} \odot \mathsf{i}))$
    \EndFor
\EndFunction

\Function{Revoke}{$\mathsf{master\_key}, \mathsf{preimage}$}
    \State $\mathsf{S}.\mathsf{add}(\mathsf{preimage})$
    \For{$\mathsf{i}~\textbf{in}~[0\ldots\mathsf{N}]$}
        \If{$\mathsf{S}.\mathsf{has}(H(\mathsf{master\_key} \odot \mathsf{i}))$}
            \State $\mathsf{S}.\mathsf{remove}(H(\mathsf{master\_key} \odot \mathsf{i}))$
            \State \Return
        \EndIf
    \EndFor
\EndFunction

\Function{Check}{$\mathsf{preimage}$}
    \State \Return $\mathsf{S}.\mathsf{has}(\mathsf{preimage})$
\EndFunction

\end{algorithmic}
\end{algorithm}

\subsection{Password Generation Algorithm}
\label{sec:generation}

Finally, we discuss the issue of password policy support. As services enforce a wide array of seemingly arbitrary requirements on password length, complexity, character sets, etc., the final step of converting a site-specific preimage into a generated password must be sufficiently flexible to ensure compatibility with a variety of services.

Rather than implementing an obscure password policy notation standard such as the Password Policy Markup Language (PPML) \cite{horsch2016password} or NIST 800-63-3 \cite{grassi2017draft}, we chose to support any regular password policy (i.e., any policy that can be expressed as a regular expression), as this includes all instances of the former, as well as most conceivable realistic password policies.

\begin{figure}[H]
\centering
\includegraphics[width=\linewidth]{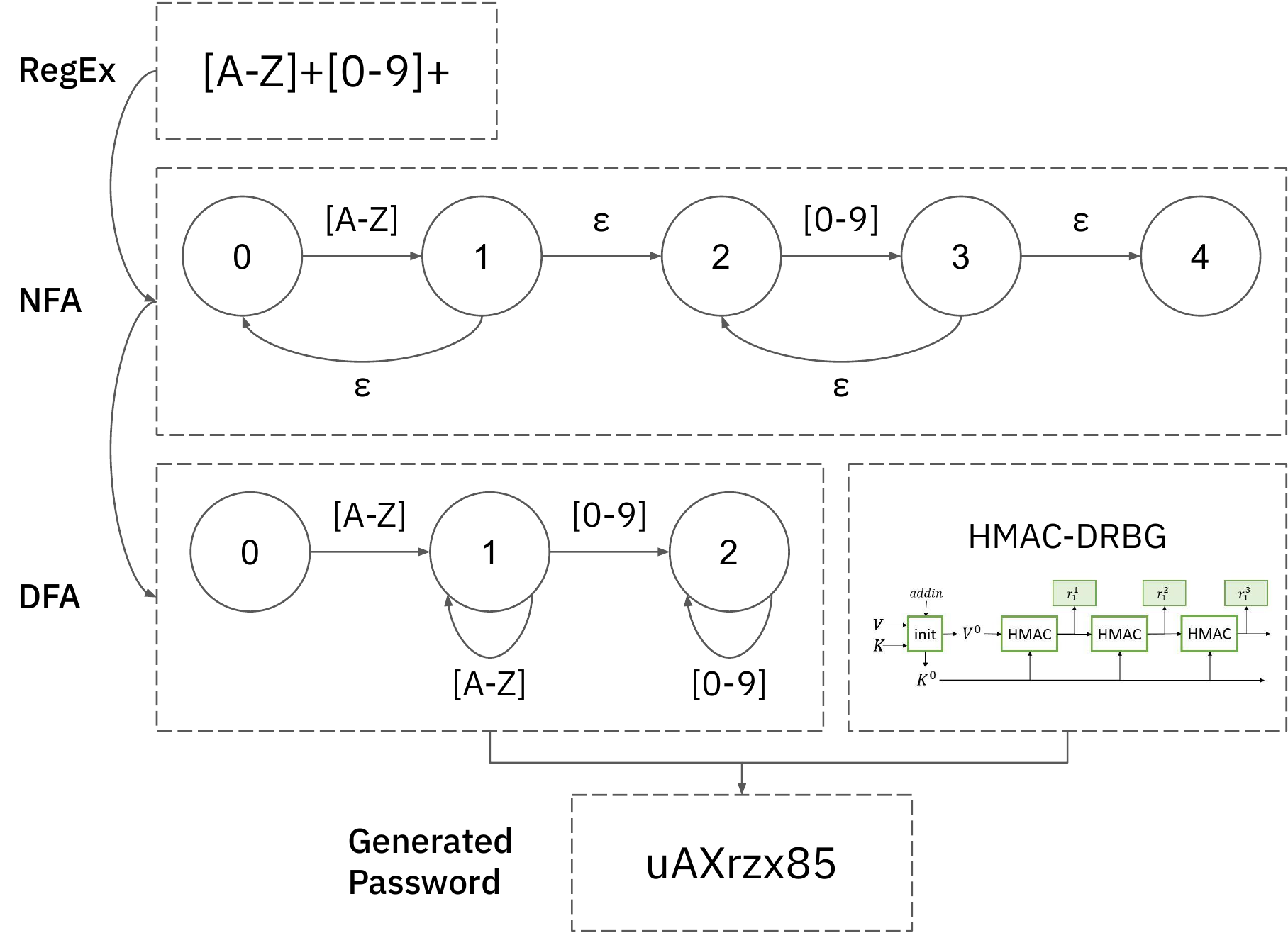}
\caption{Algorithm for generating random instances of regular password policy.}
\label{fig:regex}
\end{figure}

To support all regular password policies, we suggest using a random regular DFA traversal algorithm such as Xeger \cite{team2009xeger} as shown in Fig. \ref{fig:regex}. First, a password policy, stored as a regular expression, is converted into an NFA using the McNaughton–Yamada–Thompson algorithm \cite{thompson1968programming}. That NFA is, in turn, converted into a DFA using the subset construction algorithm \cite{rabin1959finite}. Finally, a DFA traversal algorithm, such as Xeger \cite{team2009xeger}, performs a random walk of the generated DFA. The source of randomness is overridden to use a cryptographically secure PRNG, such as HMAC-DRBG \cite{barker_recommendation_2015}, seeded using the preimage for a given service. Thus, for a given service, using the same preimage and password policy, the same password will always be generated.

The composition of these methods is shown in Alg.~\ref{alg:generation}. According to our survey of 45 existing DPGs in \S\ref{sec:survey}, we believe we are the first to propose this general approach.


\begin{algorithm}[H]
\caption{MFDPG Password Generation Algorithm}
\label{alg:generation}
\begin{algorithmic}[1]
\Require $\mathsf{G}$ is a random regex DFA traverser (e.g., Xeger \cite{team2009xeger})
\Require $\mathsf{R}$ is HMAC-DRBG per NIST SP 800-90 \cite{barker_recommendation_2015}.

\Function{Generate}{$\mathsf{preimage}, \mathsf{policy}$}
    \State \Return $\mathsf{G}(\mathsf{policy}, \mathsf{R}(\mathsf{preimage}))$
\EndFunction

\end{algorithmic}
\end{algorithm}

\section{Evaluation}
\label{sec:evaluation}

We evaluate the proposal on three grounds. First, we implement and benchmark MFDPG in a typical use case. Next, we systematically verify its compatibility with a large number of line services. Finally, we present semi-formal security arguments to demonstrate that our scheme should, in theory, satisfy the desired security properties of \S\ref{sec:problem}.

\subsection{Implementation}

To demonstrate the immediate practical utility of MFDPG and provide a blueprint for its deployment, we implemented a fully-featured open-source MFDPG JavaScript library, which is offered under a BSD license. Our implementation supports all of the previously discussed features, including multi-factor authentication, portability, revocation, and password generation based on regular expressions. During our evaluation, we interacted with MFDPG using a simple command-line interface. However, the library could also readily be used to produce a website, browser extension, mobile application, desktop program, or other means of accessing MFDPG.

\subsection{Performance}
To evaluate the performance of MFDPG in a practical setting, we benchmarked our JavaScript implementation using Node.js v16.15.0 on Windows 10 v21H2.

We used Argon2id as the underlying KDF, with p=1, t=2, and m=24576.
Our test device contains an AMD Ryzen 9 5950X (16-core, 3.4 GHz) processor and 128 GB of system memory.
However, only single-thread performance is relevant with the chosen KDF parameters, and significantly less than 1~GB of system memory is ever utilized.

We chose to set the maximum allowed revocations (labeled $\mathsf{N}$ in Alg. \ref{alg:revocation}) to 4096, and configured the underlying Cuckoo filter to have a false positive rate of 0.0001.

\eject

To benchmark the performance of MFDPG, we performed the following operations, in sequence:

\begin{enumerate}[leftmargin=*]
    \item Create a new MFDPG instance with three factors.
    \item Export the public parameters as a string.
    \item Reload the MFDPG instance using the same three factors.
    \item Generate a password for a service using a regular policy.
    \item Revoke the newly generated password.
\end{enumerate}

\begin{figure}[H]
\centering
\includegraphics[width=\linewidth]{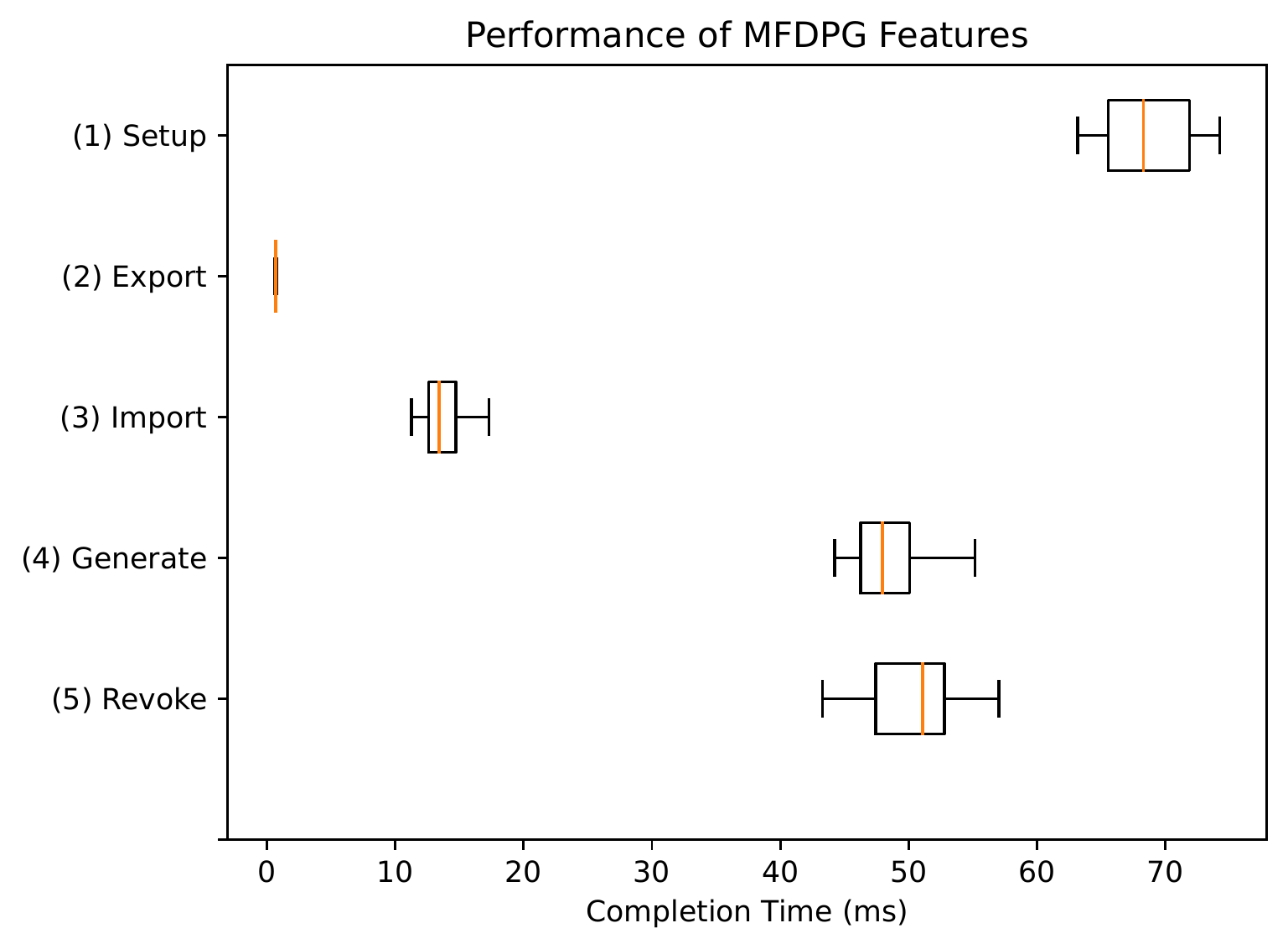}
\caption{Benchmarking results of each tested MFDPG feature (box plot).}
\label{fig:performance}
\end{figure}

This process was then repeated 100 times. The time taken to perform each step is shown as a box plot in Fig. \ref{fig:performance}. The results show that no individual operation took more than 100~ms on average. Thus, the latency of MFDPG is well within the bounds that most users will comfortably tolerate \cite{arapakis_impact_2021}.

\subsection{Compatibility}

Next, we evaluated our MFDPG implementation from the perspective of compatibility with popular websites and their respective password policies.
As the additional flexibility granted by support for all regular password policies is amongst our claimed contributions, we felt it necessary to demonstrate that this approach does indeed enhance the practicality of the resulting DPG. Thus, we tested the process of creating an account for each of the 100 most popular websites on the internet, according to their Alexa rank as of Feb 1, 2023 \cite{noauthor_alexa_nodate}. 

We found that MFDPG was indeed flexible enough to support the password policies of all 100 evaluated services. For instance, Apple's password policy does not allow three or more consecutive identical characters \cite{noauthor_apple_nodate}, a constraint that MFDPG readily handled but that no existing DPG of the 45 surveyed in \S\ref{sec:survey} could correctly enforce.

The ease of representing all encountered password policies as regular expressions suggests that a practical deployment of MFDPG could simply utilize a central database of password policies to ensure that generated passwords are automatically compliant with all popular services without requiring any manual intervention of the end user.

\eject

\subsection{Security Arguments}

We conclude our evaluation with brief arguments for why the MFDPG scheme of \S\ref{sec:approach}, as implemented herein, satisfies our original security goals in \S\ref{sec:goals}.
While a formal proof framework for multi-factor key derivation exists \cite{krawczyk, mfkdf}, there is no clear equivalent for deterministic password generation. As such, our arguments are instead based on a semi-formal reduction to the security properties of the underlying cryptographic primitives, such as MFKDF and Argon2.

\begin{enumerate}[leftmargin=*]
    \itemsep 1em
    \item \textbf{Correctness}. We first argue for the correctness (determinism) of the password generation algorithm.
    The Thompson algorithm \cite{thompson1968programming} and subset construction algorithm \cite{rabin1959finite} are both completely deterministic, i.e., they will always generate the same DFA given the same regular expression. The Xeger algorithm \cite{team2009xeger} performs a walk of the DFA, with each decision made according to the output of a PRG.
    When using HMAC-DRBG \cite{barker_recommendation_2015} as the underlying PRG, the output is deterministic when using a static HMAC key.
    Thus, given the same HMAC key (preimage) as input, Xeger always performs the same traversal of the DFA and thus generates the same password as the output string.
    
    \medskip

    Now, given the correctness of the password generation scheme, the correctness of MFDPG reduces to the ability to generate the same preimage for a given service using a fixed set of factors. This, in turn, is guaranteed by the ``correctness'' property of MFKDF and by the determinism of the underlying KDF such as Argon2.
    
    \item \textbf{Safety}. Similarly, we begin by arguing the safety (pseudorandomness) of the password generation algorithm. We note that HMAC-DBRG \cite{barker_recommendation_2015} is a cryptographically secure PRNG; it has been proven, with machine-checked proofs, that its output is indistinguishable from random \cite{ye2016notorious}. It follows that HMAC-DBRG inherits the avalanche property of HMAC: if the preimage is changed slightly (e.g, flipping a single bit), the generated bits change significantly (e.g., half the bits flip). Thus, any change to the preimage will result in a significantly different walk of the DFA, and will thus generate a significantly different password (except with negligible probability w.r.t. the policy search space).
    
    \medskip

    Given the safety of the password generation scheme, the safety of MFDPG depends on generating a different preimage for a given service when using different factors. This, in turn, is guaranteed by the ``safety'' property of MFKDF to generate a different master key if the factors change, and by the ``collision resistance'' of the underlying KDF to generate a different preimage if the master key changes.

    \item \textbf{Secretless}. We note that the MFDPG scheme does not store site-specific ciphertexts in any location. In fact, the scheme only suggests two persisted values: a Cuckoo filter for revocation, and the public parameters for dynamic MFKDF factors, such as HOTP, TOTP, and YubiKey.

    \medskip

    The former contains hashes of revoked credentials, but cannot easily be reversed to reveal the actual credentials due to the one-way nature of cryptographic hash functions. Even if somehow reversed, it would not be possible to test which service the credential belongs to without knowing the master key due to the one-way nature of the KDF used.

    \medskip

    The latter is trustless (and in fact, can be stored on a public \cite{mfkdf_wallet}), and in any case, is not site-specific.

    \medskip
    
    In our construction, the cardinality of the Cuckoo filter is static regardless of the number of revoked credentials. Thus, there are no persisted ciphertexts or other values that reveal any information about the services utilized by the user or even the number of services ever utilized or revoked.

    \item \textbf{Brute-Force Resistance}.
    First, we note that adversaries cannot directly reverse a site-specific password to obtain the master key or authentication factors, due in part to the one-way nature of the KDF, such as Argon2 \cite{argon2}.

    \medskip

    Moreover, adversaries can no longer use a site-specific password to perform a brute-force attack on the master password. Due to the ``exponential security'' of MFKDF, attackers must now simultaneously guess every possible combination of input factors, and cannot attack any input factor individually (see Nair and Song Thm. 2 \cite{mfkdf}).
    
    \item \textbf{Compatibility}. Next, we argue that MFDPG is able to generate passwords compliant with any regular password policy. First, the Thompson algorithm \cite{thompson1968programming} can convert any regular expression to an NFA, and any NFA can be converted to its equivalent DFA \cite{rabin1959finite}.
    When performing a random walk of that DFA, the Xeger \cite{team2009xeger} algorithm will always end on an ``accept'' state, and will thus always produce a string that matches the original regular expression.
    
    \item \textbf{Revocability}. When revoking a credential in MFDPG, the preimage corresponding to that credential is added to the Cuckoo filter. When generating that credential, the preimage will be found to be in the Cuckoo filter (with $p=1$), and a new preimage will be generated. As discussed in relation to the safety property, changing the preimage will result in a different password being generated (except with negligible probability).    
    Thus, revoking a password guarantees that the next password generated will be different (except with negligible probability), and MFDPG achieves this without requiring users to change their input factors or manually remember a counter value.
    
    \item \textbf{Privacy}. In implementing the above properties, MFDPG does not store any site-specific values, as discussed in relation to the ``secretless'' property. Thus, MFDPG does not store site-specific values that allow any outside adversary to identify the services utilized by its user.
\end{enumerate}
\section{Discussion}
\label{sec:discussion}

Since their introduction in 2002, deterministic password generators have been categorically rejected by consumers and researchers alike as being a viable alternative to conventional password managers, with even the most popular DPGs having a negligible market share of overall password management usage. Our survey of 45 existing DPGs shows that this decision is not entirely without merit, with current DPGs demonstrating an array of issues hindering their security, privacy, and usability.

It does not, however, follow from this that DPGs should be entirely discarded as a meritorious field of study.
In light of new cryptographic primitives, we hope to revive academic interest in DPGs as a password management approach.

While the MFDPG does not necessarily rise to the level of a production-ready system, it nevertheless constitutes a significant improvement over existing DPG proposals. By strategically combining relevant data structures, such as Cuckoo filters, algorithms, such as DFA traversal, and cryptographic primitives, such as MFKDF, we have presented a serious attempt at building a modern-day DPG, and hope to demonstrate that most of the issues surrounding DPGs are not intrinsic, and can in fact be addressed by diligent engineering.

In light of the recent massive data breach associated with LastPass \cite{toubba_notice_2022} and previous string of incidents associated with major password managers \cite{lastpass_security}, there has been an uptick in academic research and discussion around secure cryptography and system designs for password management. Due to the fundamental appeal of not storing passwords, neither in plaintext nor encrypted, in any location, we hope DPGs, such as MFDPG, remain a part of this conversation.

One final advantage of MFDPG worthy of highlighting is its ability to progressively upgrade weak password-only websites to effectively use strong MFA, simply by using an MFDPG-generated password on the client side. When configured with authentication factors like HOTP, TOTP, or YubiKeys, the master key, and thus the site-specific password, cannot be derived without using multiple strong authentication factors, even if the site itself does not accept factors other than the generated password. Moreover, MFDPG links the ability to derive the site-specific password to the correctness of these multiple factors in a strong cryptographic sense, unlike password managers that may support MFA, but ultimately encrypt credentials with a password-derived key.
This, on its own, constitutes a significant advantage of MFDPG, particularly for those who frequently interact with legacy systems with limited multi-factor authentication support, but still require trustless cryptographic enforcement of multiple strong factors. 

\subsection{Limitations}

There are a number of trade-offs associated with MFDPG, both in comparison with prior DPGs, as well as in comparison with conventional password managers.

First, unlike some DPGs, MFDPG is not completely stateless, storing both a Cuckoo filter for revocation and public parameters for certain MFKDF factors. This is a necessary concession to support multi-factor authentication and password revocation, and because these parameters require no trust assumptions, they can be stored locally or in the cloud without weakening the security of the scheme. Still, the need to store any information at all may be viewed as a drawback compared to the simplest forms of deterministic password generation.
Also, while not vulnerable to a breach of the stored materials, the system may still be vulnerable to active spyware that can obtain the master key from system memory; however, this limitation likely applies to DPGs and password managers alike.

With respect to revocation, the use of Cuckoo hashing also has potential drawbacks, being a probabilistic data structure. Namely, there is a potential for a credential revocation to incorrectly revoke another credential that is actively in use. Correctly configuring the parameters can make the probability of this small, but non-zero. Alternatively, the Cuckoo filter could be replaced with another construction that supports set addition, subtraction, and membership testing.

Furthermore, if a credential has indeed been revoked, the current construction suggests repeating the KDF invocation to create a new preimage. While this is advantageous from a security perspective, in the event that a credential has been revoked many times, generating that credential may be very slow due to the need to repeatedly invoke the adaptive KDF.

Finally, the MFDPG scheme depends on the security properties of MFKDF, and thus also inherits many of its limitations. In particular, MFKDF supports a limited set of authentication factors, and implementing new factors requires a specific, purpose-built factor constriction. While the supported factors include popular factors like HOTP, TOTP, YubiKey, etc., schemes relying on MFKDF may not be able to support as many authentication factors as a typical centralized password manager.
Further, these factors must be entered and verified simultaneously, rather than sequentially, in order to achieve the ``exponential security'' of MFKDF. While this provides a significant security advantage in resisting brute-force attacks, it may somewhat degrade the usability of the system.

Some of the limitations presented in this section are intrinsically implicated in deterministic password generation, but many can be rectified, at least in part, with further cryptographic or system design improvements over time.

\subsection{Future Work}

Currently, the generation algorithm requires a password policy to be specified as a regular expression. While these policies can be saved for popular services in a central database, the significant effort would be required to update and maintain such a database, and less popular services would inevitably be excluded. In the future, researchers could investigate an approach that automatically determines the password policy for a given service, such as through the use of language models.

The revocation component of the system could also potentially be improved. For example, some form of multi-party computation, such as a private set intersection algorithm, could perhaps be used to communicate with a central server to check whether a credential has been revoked on any device without revealing the identity of the credential being checked.

One major criticism of DPG schemes not discussed thus far is the lack of support for existing credentials. Currently, DPG systems, including MFDPG, have a high upfront adoption cost, as they essentially require a user to change every password for all of their existing services to the DPG-generated password.
Instead of this, a hybrid implementation, supporting both deterministic password generation and conventional password storage may be more user-friendly in the short term.

Finally, we hope to see usable security or HCI research that evaluates the usability of DPGs, such as MFDPG, in comparison with conventional password managers, via a controlled user study. As usability is likely as significant a factor in hindering DPG adoption as are security and privacy concerns, further advancements in this area are greatly encouraged.

\eject
\section{Conclusion}

In our brief survey covering over 20 years of DPG history, we identified a number of key issues hindering the mainstream adoption of DPGs, despite their theoretically strong advantage of not storing any site-related secrets. 
In this paper, we have chosen to view these problems as opportunities for improvement rather than disqualifying the field of DPG writ large.

MFDPG completely solves some of the major drawbacks of existing DPGs, such as password policy compatibility and multi-factor authentication support, and at a minimum, makes significant progress towards remedying other flaws, such as revocability and brute-force susceptibility. By cryptographically incorporating the entropy of multiple strong authentication factors into the password generation process, it also has the effect of turning any password-based authentication flow into a secure multi-factor authenticated login process.

In light of recent major security incidents with popular password managers, researchers are rightly taking a second look at the way we handle password management. 
While there might be a temptation to ignore DPGs as an outdated approach, we hope this work serves as a launching point for further research into secure, usable DPGs, and believe DPGs should remain an important part of this conversation moving forward.

\section*{Availability}
Our fully-functional JavaScript implementation of MFDPG, with support for multi-factor authentication, portability, revocation, and password generation based on regular expressions, is available here under a BSD license:\\

\centerline{\url{https://github.com/multifactor/mfdpg}}

\section*{Acknowledgments}
We appreciate the advice of Conor Gilsenan.
This work was supported by the National Science Foundation, the National Physical Science Consortium, the Fannie and John Hertz Foundation, and the Center for Responsible, Decentralized Intelligence. Any opinions, findings, and recommendations expressed in this material are those of the authors and do not necessarily reflect the views of the supporting entities.

\bibliographystyle{IEEEtranS}
\bibliography{references}

\begin{thebibliography}{10}
\providecommand{\url}[1]{#1}
\csname url@samestyle\endcsname
\providecommand{\newblock}{\relax}
\providecommand{\bibinfo}[2]{#2}
\providecommand{\BIBentrySTDinterwordspacing}{\spaceskip=0pt\relax}
\providecommand{\BIBentryALTinterwordstretchfactor}{4}
\providecommand{\BIBentryALTinterwordspacing}{\spaceskip=\fontdimen2\font plus
\BIBentryALTinterwordstretchfactor\fontdimen3\font minus
  \fontdimen4\font\relax}
\providecommand{\BIBforeignlanguage}[2]{{%
\expandafter\ifx\csname l@#1\endcsname\relax
\typeout{** WARNING: IEEEtranS.bst: No hyphenation pattern has been}%
\typeout{** loaded for the language `#1'. Using the pattern for}%
\typeout{** the default language instead.}%
\else
\language=\csname l@#1\endcsname
\fi
#2}}
\providecommand{\BIBdecl}{\relax}
\BIBdecl

\bibitem{1password_arch}
\BIBentryALTinterwordspacing
1Password, ``\BIBforeignlanguage{en}{{1Password} {Security} {Design}},'' 2021.
  [Online]. Available:
  \url{https://1passwordstatic.com/files/security/1password-white-paper.pdf}
\BIBentrySTDinterwordspacing

\bibitem{credential_stuffing}
\BIBentryALTinterwordspacing
Akamai, ``2020 state of the internet.'' [Online]. Available:
  \url{https://www.akamai.com/site/en/documents/state-of-the-internet/soti-security-credential-stuffing-in-the-media-industry-report-2020.pdf}
\BIBentrySTDinterwordspacing

\bibitem{noauthor_alexa_nodate}
\BIBentryALTinterwordspacing
``Alexa {Top} {Websites} {\textbar} {Last} {Save} before it was closed.''
  [Online]. Available: \url{https://www.expireddomains.net/alexa-top-websites/}
\BIBentrySTDinterwordspacing

\bibitem{noauthor_apple_nodate}
\BIBentryALTinterwordspacing
``\BIBforeignlanguage{en}{Apple - {Dumb} {Password} {Rules}}.'' [Online].
  Available: \url{https://dumbpasswordrules.com/sites/apple/}
\BIBentrySTDinterwordspacing

\bibitem{Aprico}
\BIBentryALTinterwordspacing
``Aprico.'' [Online]. Available:
  \url{https://addons.mozilla.org/en-US/firefox/addon/aprico-free-password-manager//}
\BIBentrySTDinterwordspacing

\bibitem{arapakis_impact_2021}
\BIBentryALTinterwordspacing
I.~Arapakis, S.~Park, and M.~Pielot, ``Impact of {Response} {Latency} on {User}
  {Behaviour} in {Mobile} {Web} {Search},'' in \emph{Proceedings of the 2021
  {Conference} on {Human} {Information} {Interaction} and {Retrieval}}, Mar.
  2021, pp. 279--283, arXiv:2101.09086 [cs]. [Online]. Available:
  \url{http://arxiv.org/abs/2101.09086}
\BIBentrySTDinterwordspacing

\bibitem{noauthor_4_2016}
\BIBentryALTinterwordspacing
T.~Arcieri, ``\BIBforeignlanguage{en}{4 fatal flaws in deterministic password
  managers},'' Nov. 2016. [Online]. Available:
  \url{http://tonyarcieri.com/4-fatal-flaws-in-deterministic-password-managers}
\BIBentrySTDinterwordspacing

\bibitem{Art}
\BIBentryALTinterwordspacing
``Art.'' [Online]. Available:
  \url{https://addons.mozilla.org/en-US/firefox/addon/art-password/}
\BIBentrySTDinterwordspacing

\bibitem{barker_recommendation_2015}
\BIBentryALTinterwordspacing
E.~Barker and J.~Kelsey, ``\BIBforeignlanguage{en}{Recommendation for {Random}
  {Number} {Generation} {Using} {Deterministic} {Random} {Bit} {Generators}},''
  National Institute of Standards and Technology, Tech. Rep. NIST Special
  Publication (SP) 800-90A Rev. 1, Jun. 2015. [Online]. Available:
  \url{https://csrc.nist.gov/publications/detail/sp/800-90a/rev-1/final}
\BIBentrySTDinterwordspacing

\bibitem{argon2}
A.~Biryukov, D.~Dinu, and D.~Khovratovich, ``Argon2: New generation of
  memory-hard functions for password hashing and other applications,'' in
  \emph{IEEE EuroS\&P}, 2016, pp. 292--302.

\bibitem{bitwarden}
\BIBentryALTinterwordspacing
``Bitwarden open source password manager | bitwarden.'' [Online]. Available:
  \url{https://bitwarden.com}
\BIBentrySTDinterwordspacing

\bibitem{BPasswd}
\BIBentryALTinterwordspacing
``Bpasswd.'' [Online]. Available:
  \url{https://web.archive.org/web/20181102000510/https://addons.mozilla.org/en-US/firefox/addon/bpasswd/}
\BIBentrySTDinterwordspacing

\bibitem{BPasswd2}
\BIBentryALTinterwordspacing
``Bpasswd2.'' [Online]. Available:
  \url{https://web.archive.org/web/20181102202022/https://addons.mozilla.org/en-US/firefox/addon/bpasswd2//}
\BIBentrySTDinterwordspacing

\bibitem{CCTOO}
\BIBentryALTinterwordspacing
``Cctoo.'' [Online]. Available:
  \url{https://addons.mozilla.org/en-GB/firefox/addon/cctoo/}
\BIBentrySTDinterwordspacing

\bibitem{dashlane_arch}
\BIBentryALTinterwordspacing
Dashlane, ``Security white paper,'' Mar. 2021. [Online]. Available:
  \url{https://www.dashlane.com/download/whitepaper-en.pdf}
\BIBentrySTDinterwordspacing

\bibitem{determ-pwgen}
\BIBentryALTinterwordspacing
``determ-pwgen.'' [Online]. Available:
  \url{https://github.com/I3ck/determ-pwgen}
\BIBentrySTDinterwordspacing

\bibitem{Domain}
\BIBentryALTinterwordspacing
``Domain.'' [Online]. Available:
  \url{https://web.archive.org/web/20181102054736/https://addons.mozilla.org/en-US/firefox/addon/domain-password-generator/}
\BIBentrySTDinterwordspacing

\bibitem{Extrasafe}
\BIBentryALTinterwordspacing
``Extrasafe.'' [Online]. Available:
  \url{https://web.archive.org/web/20181102205616/https://addons.mozilla.org/en-US/firefox/addon/extrasafe/}
\BIBentrySTDinterwordspacing

\bibitem{fan2014cuckoo}
B.~Fan, D.~G. Andersen, M.~Kaminsky, and M.~D. Mitzenmacher, ``Cuckoo filter:
  Practically better than bloom,'' in \emph{Proceedings of the 10th ACM
  International on Conference on emerging Networking Experiments and
  Technologies}, 2014, pp. 75--88.

\bibitem{florencio_large_2006}
\BIBentryALTinterwordspacing
D.~Florencio and C.~Herley, ``A {Large} {Scale} {Study} of {Web} {Password}
  {Habits},'' Microsoft, Tech. Rep. MSR-TR-2006-166, Nov. 2006. [Online].
  Available:
  \url{https://www.microsoft.com/en-us/research/publication/a-large-scale-study-of-web-password-habits/}
\BIBentrySTDinterwordspacing

\bibitem{grassi2017draft}
P.~A. Grassi, M.~E. Garcia, and J.~L. Fenton, ``Draft nist special publication
  800-63-3 digital identity guidelines,'' \emph{National Institute of Standards
  and Technology, Los Altos, CA}, 2017.

\bibitem{password_reuse}
A.~Hanamsagar, S.~S. Woo, C.~Kanich, and J.~Mirkovic,
  ``\BIBforeignlanguage{en}{How {Users} {Choose} and {Reuse} {Passwords}},''
  p.~16.

\bibitem{rfc6234}
\BIBentryALTinterwordspacing
T.~Hansen and D.~E. Eastlake~3rd, ``{US} {Secure} {Hash} {Algorithms} ({SHA}
  and {SHA}-based {HMAC} and {HKDF}),'' Internet Engineering Task Force,
  Request for {Comments} RFC 6234, May 2011. [Online]. Available:
  \url{https://datatracker.ietf.org/doc/rfc6234}
\BIBentrySTDinterwordspacing

\bibitem{Hash_Password}
\BIBentryALTinterwordspacing
``Hash password.'' [Online]. Available:
  \url{https://web.archive.org/web/20181102030655/https://addons.mozilla.org/en-US/firefox/addon/hash-password-generator/}
\BIBentrySTDinterwordspacing

\bibitem{hash0}
\BIBentryALTinterwordspacing
``hash0.'' [Online]. Available:
  \url{https://web.archive.org/web/20181103000530/https://addons.mozilla.org/en-US/firefox/addon/hash0/}
\BIBentrySTDinterwordspacing

\bibitem{Hasher_Plus}
\BIBentryALTinterwordspacing
``Hasher plus.'' [Online]. Available:
  \url{https://addons.mozilla.org/en-US/firefox/addon/password-hasher-plus/}
\BIBentrySTDinterwordspacing

\bibitem{HashPass}
\BIBentryALTinterwordspacing
``Hashpass.'' [Online]. Available:
  \url{https://web.archive.org/web/20181101222025/https://addons.mozilla.org/en-US/firefox/addon/hashpass-firefox/}
\BIBentrySTDinterwordspacing

\bibitem{horsch2016password}
M.~Horsch, M.~Schlipf, S.~Haas, J.~Braun, and J.~Buchmann, ``Password policy
  markup language,'' 2016.

\bibitem{ooba}
\BIBentryALTinterwordspacing
P.~Identity, ``\BIBforeignlanguage{en}{What is {Out}-of-{Band} {Authentication}
  ({OOBA})?}'' [Online]. Available:
  \url{https://www.pingidentity.com/en/resources/blog/post/what-is-out-of-band-authentication-ooba.html}
\BIBentrySTDinterwordspacing

\bibitem{rfc2898}
\BIBentryALTinterwordspacing
B.~Kaliski, ``{PKCS} \#5: {Password}-{Based} {Cryptography} {Specification}
  {Version} 2.0,'' Internet Engineering Task Force, Request for {Comments} RFC
  2898, Sep. 2000, num Pages: 34. [Online]. Available:
  \url{https://datatracker.ietf.org/doc/rfc2898}
\BIBentrySTDinterwordspacing

\bibitem{karp2002site}
A.~H. Karp, ``Site-specific passwords,'' \emph{HP Labs Technical Report}, 2002.

\bibitem{keepassx}
\BIBentryALTinterwordspacing
``Keepassx.'' [Online]. Available: \url{https://www.keepassx.org/}
\BIBentrySTDinterwordspacing

\bibitem{krawczyk}
\BIBentryALTinterwordspacing
H.~Krawczyk, ``Cryptographic {Extraction} and {Key} {Derivation}: {The} {HKDF}
  {Scheme}.'' [Online]. Available: \url{https://eprint.iacr.org/2010/264}
\BIBentrySTDinterwordspacing

\bibitem{lastpass_arch}
\BIBentryALTinterwordspacing
LastPass, ``Our {Zero}-{Knowledge} {Security} {Model}.'' [Online]. Available:
  \url{https://www.lastpass.com/security/zero-knowledge-security}
\BIBentrySTDinterwordspacing

\bibitem{LastWord}
\BIBentryALTinterwordspacing
``Lastword.'' [Online]. Available: \url{https://github.com/LoneFry/LastWord}
\BIBentrySTDinterwordspacing

\bibitem{Magic}
\BIBentryALTinterwordspacing
``Magic.'' [Online]. Available:
  \url{https://github.com/arantius/magic-password-generator}
\BIBentrySTDinterwordspacing

\bibitem{masterpassword}
\BIBentryALTinterwordspacing
``masterpassword.'' [Online]. Available:
  \url{https://addons.mozilla.org/en-GB/firefox/addon/masterpassword-firefox/}
\BIBentrySTDinterwordspacing

\bibitem{MasterPassX}
\BIBentryALTinterwordspacing
``Masterpassx.'' [Online]. Available:
  \url{https://chrome.google.com/webstore/detail/masterpassx/acocljodaoecblhjggkadfhnbjcfgbbb}
\BIBentrySTDinterwordspacing

\bibitem{MS}
\BIBentryALTinterwordspacing
``Ms.'' [Online]. Available:
  \url{https://web.archive.org/web/20171013225154/https://addons.mozilla.org/en-US/firefox/addon/strong-password-generator//}
\BIBentrySTDinterwordspacing

\bibitem{My_Password}
\BIBentryALTinterwordspacing
``My password.'' [Online]. Available:
  \url{https://web.archive.org/web/20181101221606/https://addons.mozilla.org/en-US/firefox/addon/my-password/}
\BIBentrySTDinterwordspacing

\bibitem{mypass}
\BIBentryALTinterwordspacing
``mypass.'' [Online]. Available: \url{https://github.com/anoma-/mypass}
\BIBentrySTDinterwordspacing

\bibitem{mfkdf_wallet}
V.~Nair and D.~Song, ``Decentralizing custodial wallets with mfkdf,'' 2023.

\bibitem{mfkdf}
------, ``Multi-factor key derivation function (mfkdf) for fast, flexible,
  secure, \& practical key management,'' \emph{USENIX Security '23}, 2023.

\bibitem{fips197}
NIST, ``\BIBforeignlanguage{en}{{FIPS} 197, {Advanced} {Encryption} {Standard}
  ({AES})},'' p.~51.

\bibitem{lastpass_security}
\BIBentryALTinterwordspacing
K.~Norton, ``\BIBforeignlanguage{en-US}{{LastPass}: {Is} it a {Safe} {Password}
  {Manager}?}'' Jul. 2021. [Online]. Available:
  \url{https://www.esecurityplanet.com/products/lastpass-review/}
\BIBentrySTDinterwordspacing

\bibitem{palant_security_2016}
\BIBentryALTinterwordspacing
W.~Palant, ``\BIBforeignlanguage{en-us}{Security considerations for password
  generators},'' Apr. 2016, section: articles. [Online]. Available:
  \url{https://palant.info/2016/04/20/security-considerations-for-password-generators/}
\BIBentrySTDinterwordspacing

\bibitem{Passcrambler}
\BIBentryALTinterwordspacing
``Passcrambler.'' [Online]. Available:
  \url{https://github.com/hasherezade/password_scrambler}
\BIBentrySTDinterwordspacing

\bibitem{Passera}
\BIBentryALTinterwordspacing
``Passera.'' [Online]. Available:
  \url{https://web.archive.org/web/20181102003104/https://addons.mozilla.org/en-US/firefox/addon/passera/}
\BIBentrySTDinterwordspacing

\bibitem{Password_Hasher}
\BIBentryALTinterwordspacing
``Password hasher.'' [Online]. Available:
  \url{https://web.archive.org/web/20181002124942/https://addons.mozilla.org/en-US/firefox/addon/password-hasher/}
\BIBentrySTDinterwordspacing

\bibitem{Password_Maker_X}
\BIBentryALTinterwordspacing
``Password maker x.'' [Online]. Available:
  \url{https://web.archive.org/web/20181102214310/https://addons.mozilla.org/en-US/android/addon/password-maker-x/}
\BIBentrySTDinterwordspacing

\bibitem{PasswordBuilder}
\BIBentryALTinterwordspacing
``Passwordbuilder.'' [Online]. Available:
  \url{https://addons.mozilla.org/en-GB/firefox/addon/uniquepasswordbuilder-addon/}
\BIBentrySTDinterwordspacing

\bibitem{passwordgen}
\BIBentryALTinterwordspacing
``passwordgen.'' [Online]. Available:
  \url{https://addons.mozilla.org/en-US/firefox/addon/passwordgen-for-firefox-1/}
\BIBentrySTDinterwordspacing

\bibitem{PasswordMaker}
\BIBentryALTinterwordspacing
``Passwordmaker.'' [Online]. Available:
  \url{https://web.archive.org/web/20181002125051/https://addons.mozilla.org/en-US/firefox/addon/passwordmaker/}
\BIBentrySTDinterwordspacing

\bibitem{PasswordProtect}
\BIBentryALTinterwordspacing
``Passwordprotect.'' [Online]. Available:
  \url{https://web.archive.org/web/20181102194105/https://addons.mozilla.org/en-US/firefox/addon/passwordprotect/}
\BIBentrySTDinterwordspacing

\bibitem{PasswordShaker}
\BIBentryALTinterwordspacing
``Passwordshaker.'' [Online]. Available:
  \url{https://addons.mozilla.org/en-US/firefox/addon/passwordshaker/}
\BIBentrySTDinterwordspacing

\bibitem{pastor}
\BIBentryALTinterwordspacing
``pastor.'' [Online]. Available: \url{https://github.com/appnician/pastor}
\BIBentrySTDinterwordspacing

\bibitem{Pegasus}
\BIBentryALTinterwordspacing
``Pegasus.'' [Online]. Available:
  \url{https://addons.mozilla.org/en-US/firefox/addon/pegasus-password-generator/}
\BIBentrySTDinterwordspacing

\bibitem{scrypt}
\BIBentryALTinterwordspacing
C.~Percival, ``\BIBforeignlanguage{en}{{Stronger} {Key} {Derivation} {via}
  {Sequential} {Memory}-{Hard} {Functions}},'' p.~16. [Online]. Available:
  \url{https://www.researchgate.net/publication/252853607_Stronger_key_derivation_via_sequential_memory-hard_functions}
\BIBentrySTDinterwordspacing

\bibitem{Persistent_Generator}
\BIBentryALTinterwordspacing
``Persistent generator.'' [Online]. Available:
  \url{https://web.archive.org/web/20181101232611/https://addons.mozilla.org/en-US/firefox/addon/sebres-pwd-hasher//}
\BIBentrySTDinterwordspacing

\bibitem{Phashword}
\BIBentryALTinterwordspacing
``Phashword.'' [Online]. Available:
  \url{https://web.archive.org/web/20181102233200/https://addons.mozilla.org/en-US/firefox/addon/phashword/}
\BIBentrySTDinterwordspacing

\bibitem{bcrypt}
\BIBentryALTinterwordspacing
N.~Provos and D.~Mazi{\`e}res, ``A {Future-Adaptable} password scheme,'' in
  \emph{1999 USENIX Annual Technical Conference (USENIX ATC 99)}.\hskip 1em
  plus 0.5em minus 0.4em\relax Monterey, CA: USENIX Association, Jun. 1999.
  [Online]. Available:
  \url{https://www.usenix.org/conference/1999-usenix-annual-technical-conference/future-adaptable-password-scheme}
\BIBentrySTDinterwordspacing

\bibitem{PswGen_Toolbar}
\BIBentryALTinterwordspacing
``Pswgen toolbar.'' [Online]. Available:
  \url{https://web.archive.org/web/20181102012502/https://addons.mozilla.org/en-US/firefox/addon/pswgen-toolbar/}
\BIBentrySTDinterwordspacing

\bibitem{PwdHash}
\BIBentryALTinterwordspacing
``Pwdhash.'' [Online]. Available:
  \url{https://addons.mozilla.org/en-US/firefox/addon/pwdhash/}
\BIBentrySTDinterwordspacing

\bibitem{pwgen}
\BIBentryALTinterwordspacing
``pwgen.'' [Online]. Available: \url{https://github.com/quintopia/pwgen}
\BIBentrySTDinterwordspacing

\bibitem{python-dpg}
\BIBentryALTinterwordspacing
``python-dpg.'' [Online]. Available:
  \url{https://github.com/psmiraglia/python-dpg}
\BIBentrySTDinterwordspacing

\bibitem{rabin1959finite}
M.~O. Rabin and D.~Scott, ``Finite automata and their decision problems,''
  \emph{IBM journal of research and development}, vol.~3, no.~2, pp. 114--125,
  1959.

\bibitem{spraying}
\BIBentryALTinterwordspacing
R.~Ranjan, ``\BIBforeignlanguage{en}{Password {Spraying} {Attack} {\textbar}
  {OWASP} {Foundation}}.'' [Online]. Available:
  \url{https://owasp.org/www-community/attacks/Password_Spraying_Attack}
\BIBentrySTDinterwordspacing

\bibitem{Recall_my_password}
\BIBentryALTinterwordspacing
``Recall my password.'' [Online]. Available:
  \url{https://web.archive.org/web/20181002161104/https://addons.mozilla.org/en-US/firefox/addon/recall-my-password}
\BIBentrySTDinterwordspacing

\bibitem{reichl_keepass_nodate}
\BIBentryALTinterwordspacing
D.~Reichl, ``\BIBforeignlanguage{en}{{KeePass} {Password} {Safe}}.'' [Online].
  Available: \url{https://keepass.info/}
\BIBentrySTDinterwordspacing

\bibitem{RndPhrase}
\BIBentryALTinterwordspacing
``Rndphrase.'' [Online]. Available:
  \url{https://web.archive.org/web/20181102013645/https://addons.mozilla.org/en-US/firefox/addon/rndphrase/}
\BIBentrySTDinterwordspacing

\bibitem{ross2005stronger}
B.~Ross, C.~Jackson, N.~Miyake, D.~Boneh, and J.~C. Mitchell, ``Stronger
  password authentication using browser extensions.'' in \emph{USENIX Security
  Symposium}, vol.~17.\hskip 1em plus 0.5em minus 0.4em\relax Baltimore, MD,
  USA, 2005, p.~32.

\bibitem{SecPassGen}
\BIBentryALTinterwordspacing
``Secpassgen.'' [Online]. Available:
  \url{https://web.archive.org/web/20181103002640/https://addons.mozilla.org/en-US/firefox/addon/secpassgen//}
\BIBentrySTDinterwordspacing

\bibitem{strongpass}
\BIBentryALTinterwordspacing
``strongpass.'' [Online]. Available: \url{https://github.com/grempe/strongpass}
\BIBentrySTDinterwordspacing

\bibitem{team2009xeger}
X.~Team, ``Xeger string generator,'' 2009.

\bibitem{thompson1968programming}
K.~Thompson, ``Programming techniques: Regular expression search algorithm,''
  \emph{Communications of the ACM}, vol.~11, no.~6, pp. 419--422, 1968.

\bibitem{toubba_notice_2022}
\BIBentryALTinterwordspacing
K.~Toubba, ``Notice of {Recent} {Security} {Incident},'' Dec. 2022. [Online].
  Available:
  \url{https://blog.lastpass.com/2022/12/notice-of-recent-security-incident/}
\BIBentrySTDinterwordspacing

\bibitem{Tresor}
\BIBentryALTinterwordspacing
``Tresor.'' [Online]. Available: \url{https://github.com/mstum/TresorLib/}
\BIBentrySTDinterwordspacing

\bibitem{UniPass}
\BIBentryALTinterwordspacing
``Unipass.'' [Online]. Available:
  \url{https://web.archive.org/web/20181103011439/https://addons.mozilla.org/en-US/firefox/addon/unipass/}
\BIBentrySTDinterwordspacing

\bibitem{uPassword}
\BIBentryALTinterwordspacing
``upassword.'' [Online]. Available:
  \url{https://web.archive.org/web/20181102223643/https://addons.mozilla.org/en-US/firefox/addon/upassword/}
\BIBentrySTDinterwordspacing

\bibitem{Vault}
\BIBentryALTinterwordspacing
``Vault.'' [Online]. Available:
  \url{https://web.archive.org/web/20181102032353/https://addons.mozilla.org/en-US/firefox/addon/vault/}
\BIBentrySTDinterwordspacing

\bibitem{rfc4226}
\BIBentryALTinterwordspacing
M.~View, D.~M'Raihi, F.~Hoornaert, D.~Naccache, M.~Bellare, and O.~Ranen,
  ``{HOTP}: {An} {HMAC}-{Based} {One}-{Time} {Password} {Algorithm},'' Internet
  Engineering Task Force, Request for {Comments} RFC 4226, Dec. 2005, num
  Pages: 37. [Online]. Available:
  \url{https://datatracker.ietf.org/doc/rfc4226}
\BIBentrySTDinterwordspacing

\bibitem{rfc6238}
\BIBentryALTinterwordspacing
M.~View, J.~Rydell, M.~Pei, and S.~Machani, ``{TOTP}: {Time}-{Based}
  {One}-{Time} {Password} {Algorithm},'' Internet Engineering Task Force,
  Request for {Comments} RFC 6238, May 2011. [Online]. Available:
  \url{https://datatracker.ietf.org/doc/rfc6238}
\BIBentrySTDinterwordspacing

\bibitem{vPass}
\BIBentryALTinterwordspacing
``vpass.'' [Online]. Available:
  \url{https://web.archive.org/web/20181102055008/https://addons.mozilla.org/en-US/firefox/addon/vpass-password-generator/}
\BIBentrySTDinterwordspacing

\bibitem{ye2016notorious}
K.~Ye, ``The notorious prg: Formal verification of the hmac-drbg pseudorandom
  number generator,'' 2016.

\bibitem{yubikey}
\BIBentryALTinterwordspacing
Yubico, ``\BIBforeignlanguage{en-US}{Yubikey: Strong two-factor
  authentication}.'' [Online]. Available: \url{https://www.yubico.com/}
\BIBentrySTDinterwordspacing

\end{thebibliography}

\end{document}